\documentclass[reprint,superscriptaddress,amsmath,amssymb,aps,floatfix]{revtex4-2}
\usepackage{graphicx}
\usepackage{color}

\usepackage{dcolumn}
\graphicspath{{./Figures/}}
\usepackage{upgreek}
\usepackage{gensymb}
\usepackage{float}
\usepackage{bm}
\usepackage{xcolor}
\usepackage{physics}
\usepackage{longtable}
\usepackage{comment}

\usepackage{xr}
\makeatletter
\newcommand*{\addFileDependency}[1]{
	\typeout{(#1)}
	\@addtofilelist{#1}
	\IfFileExists{#1}{}{\typeout{No file #1.}}
}
\makeatother

\newcommand*{\myexternaldocument}[1]{
	\externaldocument{#1}
	\addFileDependency{#1.tex}
	\addFileDependency{#1.aux}
}

\myexternaldocument{ME_CFB_ver1_supps}

\begin{document}
	
\title{Large surface acoustic wave nonreciprocity in synthetic antiferromagnets}

\author{Hiroki Matsumoto}
\affiliation{Department of Physics, The University of Tokyo, Tokyo 113-0033, Japan}

\author{Takuya Kawada}
\affiliation{Department of Physics, The University of Tokyo, Tokyo 113-0033, Japan}

\author{Mio Ishibashi}
\affiliation{Department of Physics, The University of Tokyo, Tokyo 113-0033, Japan}

\author{Masashi Kawaguchi}
\affiliation{Department of Physics, The University of Tokyo, Tokyo 113-0033, Japan}

\author{Masamitsu Hayashi}
\affiliation{Department of Physics, The University of Tokyo, Tokyo 113-0033, Japan}
\affiliation{Trans-scale quantum science institute (TSQS), The University of Tokyo, Tokyo 113-0033, Japan}

\newif\iffigure
\figurefalse
\figuretrue

\date{\today}

\begin{abstract}
We have studied the transmission of surface acoustic waves (SAWs) in ferromagnetic/non-magnetic/ferromagnetic tryilayers.
The SAW scattering matrix is studied for devices with various non-magnetic spacer thickness, which defines the strength of the interlayer exchange coupling.
We find the SAW transmission amplitude depends on their propagation direction when the two ferromagnetic layers are coupled antiferromagnetically.
The degree of such SAW nonreciprocity increases with increasing exchange coupling strength and reaches 37 dB/mm for a device with the thinnest spacer layer.
These results show the potential of interlayer exchange coupled synthetic antiferromagnets for viable acoustic nonreciprocal transmission devices, such as circulators and isolators.
\end{abstract}

\maketitle

\par
Surface acoustic waves (SAWs) have played an important role in radiowave- and microwave-frequency filters and signal processing devices\cite{delsing2019jpd}. 
The high energy density and long coherence time of SAWs are particularly attractive for device applications.
Recent studies suggested and demonstrated that SAWs can be used as quanta for applications in hybrid quantum systems.
SAWs coupled to other quanta, such as superconducting qubits\cite{satzinger2018nature} and magnons\cite{hatanaka2022prap}, allow exchange of information among the quanta.
Such coupling is gaining a significant interest for applications in quantum technologies in recent years. 

Nonreciprocal propagation of SAWs in ferromagnetic thin films, a signature of SAW-magnon coupling\cite{weiler2011prl,dreher2012prb}, has been studied in various systems.
The nonreciprocity may arise in single layer ferromagnetic thin films due to helicity mismatch between the SAWs and spin waves\cite{sasaki2017prb,tateno2020prap,kuss2020prl}, the anisotropic dispersion relation of spin waves\cite{camley1987surfscirep,hernandez-minguez2020prap,babu2021nanolett} and magneto-rotation coupling\cite{xu2020sciadv}.
Nonreciprocal transmission of SAWs has also been reported in antiferromagnetically coupled ferromagnetic metal (FM)/non-magnetic metal (NM)/FM trilayers, in which one of the largest SAW nonreciprocity (per unit delay line length) was found \cite{shah2020sciadv,kuss2021prap}.
For such structure, the large non-recirpocity has been attributed to the nonreciprocal propagation of spin waves due to the dipolar-coupling between the two FM layers\cite{di2015scirep,gladii2016prb,gallardo2019prap,ishibashi2020sciadv}.

As large SAW nonreciprocity in a wide frequency range has been predicted in interlayer exchange coupled synthetic antiferromagnetic (SAF) systems\cite{verba2019prap}, here we explore its potential.
Contrary to the dipolar-coupled systems, the coupling strength of the ferromagnetic layers in SAF can be tuned by the spacer layer thickness in a significant way\cite{parkin1990prl}.
We use CoFeB/Ru/CoFeB trilayers, a prototype SAF structure, to study the transmission characteristics of the SAW.
The nonreciprocity of the SAW transmission amplitude across the trilayer is studied as a function of the spacer layer (Ru) thickness.

\par
\begin{figure}[bt]
	\begin{minipage}{1\hsize}
		\includegraphics[scale=0.3]{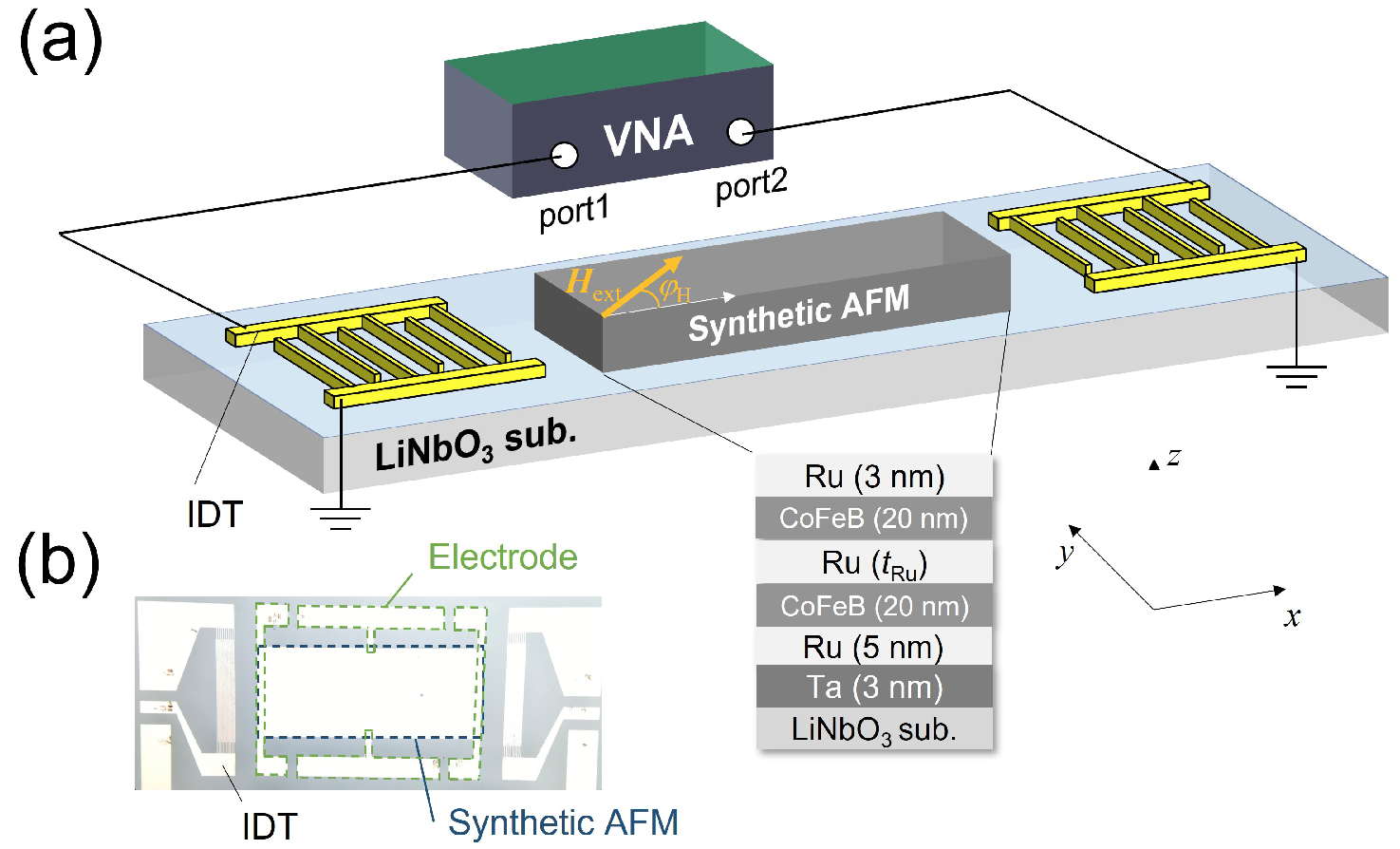}
	\end{minipage}
	\begin{tabular}{cc}
		\begin{minipage}{0.5\hsize}
			\includegraphics[scale=0.28]{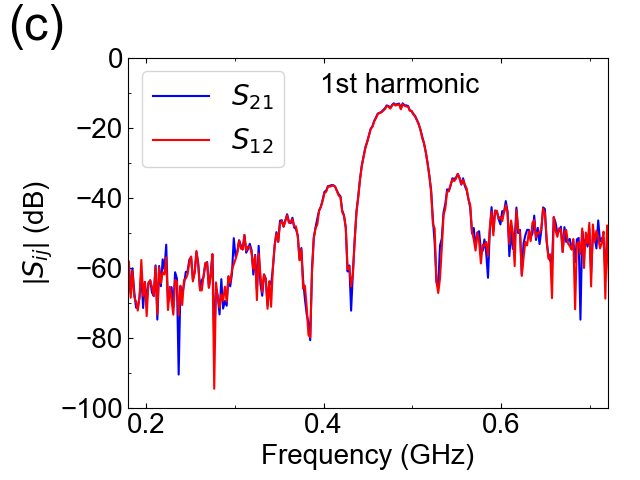}
		\end{minipage}
		\begin{minipage}{0.5\hsize}
			\includegraphics[scale=0.28]{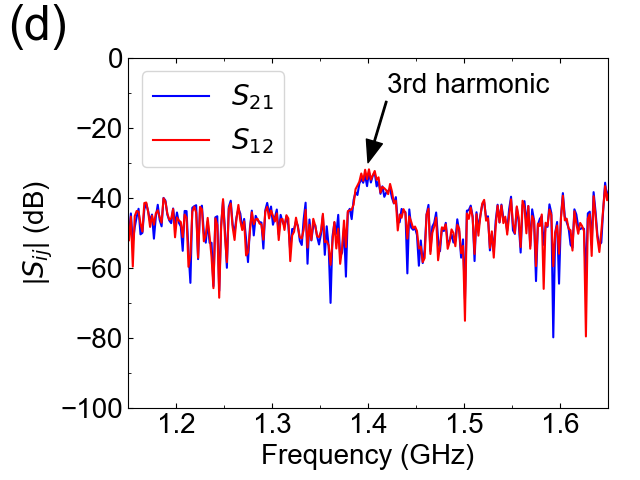}
		\end{minipage}
	\end{tabular}

	\caption{(a) Schematic illustration of the experiment setup, the film structure and the coordinate axis. (b) Photograph of the SAW device under test. The rectangular element and the attached electrode are marked for clarity. (c,d) Representative transmission spectra ($S_{21}$ and $S_{12}$) of the SAW device around the 1st (c) and 3rd (d) harmonic frequencies when a film with $t_\mathrm{Ru} = 0.46$ nm is placed in the delay line. 
		\label{fig:setup}
	}
\end{figure}

Figure 1(a) shows a schematic illustration of the experimental setup and the coordinate axis. 
The device is formed using optical photolithography and a lift-off process.
We first pattern the interdigital transducers (IDTs) for SAW excitation and detection.
Two IDTs are made of Ta(5)/Cu(50)/Pt(3) (thickness in unit of nanometer) layers that are deposited on a 128${}^\mathrm{o}$Y-cut LiNbO${}_{3}$ substrate by means of rf magnetron sputtering.
The width and intervals of the IDT fingers are both set to 2 \textmu m.
Next we form the SAF structure.
The SAF is composed of Ta(3)/Ru(5)/CoFeB(20)/Ru(${t}_\mathrm{Ru}$)/CoFeB(20)/Ru(3).
Schematic of the film structure is illustrated in Fig.~\ref{fig:setup}(a).
The bottom Ta(3)/Ru(5) is a seed layer to promote smooth growth of the succeeding layers.
The top Ru(3) is a capping layer to prevent oxidation.
The thickness ($t_\mathrm{FM}$) of the FM layer (20 nm CoFeB) is chosen to maximize the SAW-spin wave coupling, which increases with increasing $t_\mathrm{FM}$ within the range of $t_\mathrm{FM} \lesssim \lambda_p$ ($\lambda_p$ is the SAW penetration depth), while keeping the interlayer exchange coupling non-negligible, which scales with 1/$t_\mathrm{FM}$.
The film is patterned into a rectangle and placed on the delay line, \textit{i.e.}, in the region between the two IDTs.
The length and width of the rectangle are 1 mm and 0.4 mm, respectively. 
Finally, we pattern the electrodes, made of Ta(5)/Cu(50)/Pt(3), and attach them to the rectangle (see Fig~\ref{fig:setup}(b)).
Devices with different ${t}_\mathrm{Ru}$, ranging from $\sim$0.46 nm to $\sim$0.77 nm, are formed and studied. 
The rectangular element (and the electrode) is used to study the transport properties (\textit{e.g.}, longitudinal and Hall resistances) of the films.
All measurements are performed at room temperature.

\begin{figure}[t]
	\begin{tabular}{cc}
		\begin{minipage}{0.5\hsize}
			\includegraphics[scale=0.28]{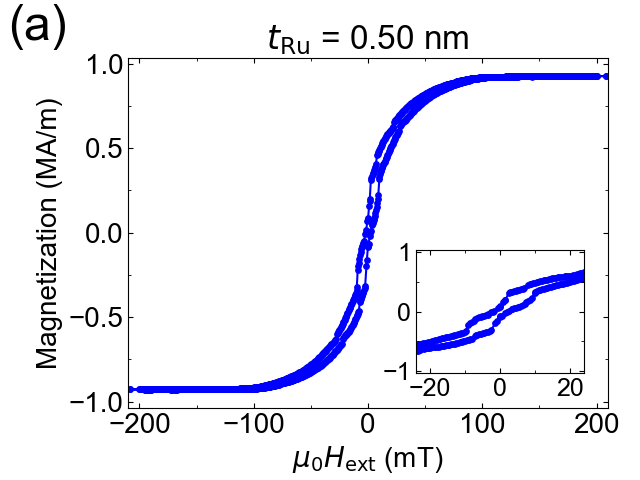}
		\end{minipage}
		\begin{minipage}{0.5\hsize}
			\includegraphics[scale=0.28]{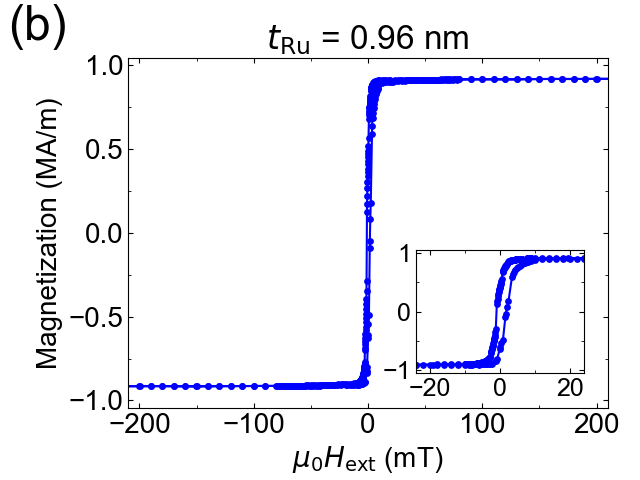}
		\end{minipage}\\
		\begin{minipage}{0.5\hsize}
			\includegraphics[scale=0.28]{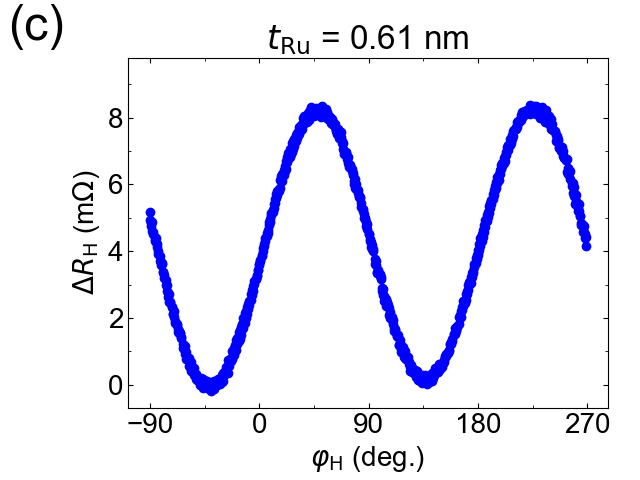}
		\end{minipage}
		\begin{minipage}{0.5\hsize}
			\includegraphics[scale=0.28]{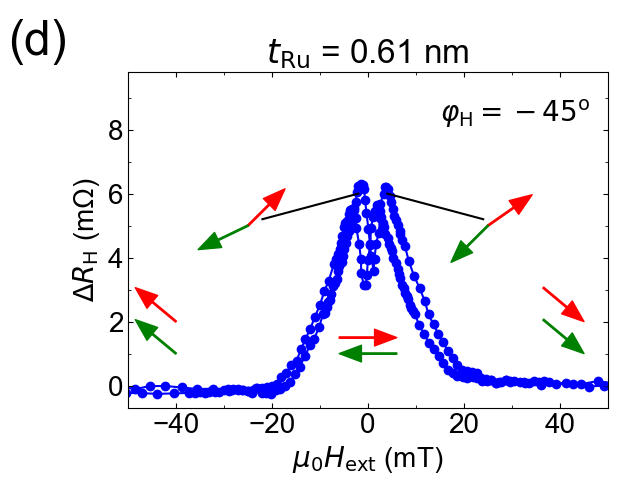}
		\end{minipage}
	\end{tabular}		
	\caption{(a,b) Magnetization-magnetic field ($M$-$H$) loops of the SAF structure deposited on Si substrates with $t_\mathrm{Ru}$ = 0.50 nm (a) and $t_\mathrm{Ru}$ = 0.96 nm (b). The insets show an expanded view near zero field. (c,d) Hall resistance $\Delta R_\mathrm{H}$ dependence on the field angle $\varphi_\mathrm{H}$ (c) and the field amplitude $\mu_{0}H_\mathrm{ext}$ (d) for a device with $t_\mathrm{Ru}$ = 0.61 nm. $|\mu_{0}H_\mathrm{ext}|$ is set to $\sim$ 60 mT for (c) and $\varphi_H$ is fixed to $\sim - 45^\mathrm{o}$ in (d). For the field angle dependence (c), we subtract contributions from the anomalous Hall effect that appears when there is an unintended misalignment of the magnetic field with the film plane. The green and red arrows in (d) illustrate the magnetization direction of the bottom and top CoFeB layer in the SAF.
		\label{fig:phe}
	}
\end{figure}

Magnetization-magnetic field ($M$-$H$) loops of the SAF are measured using vibrating sample magnetometry (VSM).
The films are deposited on $10 \times 10$ mm$^2$ Si substrates separately from the those on the LiNbO${}_{3}$ substrates.
The magnetic field is swept along the film plane.
Representative $M$-$H$ loops are shown in Figs.~\ref{fig:phe}(a) and \ref{fig:phe}(b) for films with ${t}_\mathrm{Ru} \sim 0.5$ nm and ${t}_\mathrm{Ru} \sim 1.0$ nm, respectively.
The insets show an expanded view near zero field.
As evident, the remanent magnetization at zero field is close to zero for the film with ${t}_\mathrm{Ru} \sim 0.5$ nm, whereas the film with ${t}_\mathrm{Ru} \sim 1.0$ nm shows a typical ferromagnetic square loop.
The small remanence for the former suggests the two CoFeB layers are antiferromagnetically coupled\cite{li2016scirep,ishibashi2020sciadv,waring2020prap}.
Consistent with previous studies\cite{parkin1990prl}, antiferromagnetic interlayer exchange coupling of the ferromagnetic layers is found when $t_\mathrm{Ru}$ is smaller than $\sim 1$ nm.

The magnetic response of the films are studied using transport measurements of the rectangular element.
We make use of the planar Hall effect (PHE) to study the change in the magnetization direction as the field is swept.
First, an in-plane magnetic field, sufficiently large to align the magnetization along the field, is applied to the sample.
The Hall resistance is measured while the field direction is rotated within the film plane.
We define the angle between the field and $+x$ as $\varphi_H$.
The magnitude of the external magnetic field $|\mu_0 H_\mathrm{ext}|$ is fixed to 60 mT.
Figure~\ref{fig:phe}(c) shows the $\varphi_H$ dependence of the Hall resistance ($\Delta R_\mathrm{H}$) for a device with ${t}_\mathrm{Ru}$ = 0.61 nm.
An offset Hall resistance (that includes a linear background drift, if any) is subtracted to allow comparison of the field angle (Fig.~\ref{fig:phe}(c)) and the field amplitude (Fig.~\ref{fig:phe}(d)) sweeps of the Hall resistance. 
In Fig.~\ref{fig:phe}(c), $\Delta R_\mathrm{H}$ takes a maximum (minimum) at $\varphi_H \sim 45^\mathrm{o}$, 225$^\mathrm{o}$ (-45$^\mathrm{o}$, 135$^\mathrm{o}$), a characteristic feature of the PHE\cite{mcguire1975ieee}.

Next, we sweep the field amplitude $H_\mathrm{ext}$ along the $\varphi_H \sim -45^\mathrm{o}$ direction.
Figure~\ref{fig:phe}(d) shows the corresponding $H_\mathrm{ext}$ dependence of $\Delta R_\mathrm{H}$ for the same device.
Starting from zero field, $\Delta R_\mathrm{H}$ first increases with increasing $|H_\mathrm{ext}|$.
$\Delta R_\mathrm{H}$ takes a maximum at $|\mu_0 H_\mathrm{ext}| \sim 5$ mT, above which it monotonically decreases with increasing $|H_\mathrm{ext}|$.
Above $|\mu_0 H_\mathrm{ext}| \sim 20$ mT, $\Delta R_\mathrm{H}$ saturates and exhibits a constant value.
Such changes in $\Delta R_\mathrm{H}$ with $H_\mathrm{ext}$ can be accounted for if we assume the following magnetization process takes place (see the red/green arrows depicted in Fig.~\ref{fig:phe}(d)).
At zero field, the two ferromagnetic layers form an antiparallel state owing to the interlayer exchange coupling.
Since the magnetic easy axis points along the $x$-axis, primarily defined by the shape anisotropy of the rectangular element, magnetization of the two layers, defined as $\bm{m}_1$ and $\bm{m}_2$, point along $+x$ and $-x$ at zero field. 
This is evident from the value of $\Delta R_\mathrm{H}$ at zero field, which coincides with that of $\varphi_H \sim 0^\mathrm{o}$ shown in Fig.~\ref{fig:phe}(c). 
As $|H_\mathrm{ext}|$ is increased along $\varphi_H \sim - 45^\mathrm{o}$, $\bm{m}_1$ and $\bm{m}_2$ rotate while maintaining the antiparallel state.
The increase in $\Delta R_\mathrm{H}$ when $|H_\mathrm{ext}|$ is increased from zero field is consistent with the picture that the magnetization of the two layers rotate and approach a state with $\bm{m}_1$ and $\bm{m}_2$ pointing along $\sim$45$^\mathrm{o}$ and $\sim$225$^\mathrm{o}$ from $+x$.
When $|\mu_0 H_\mathrm{ext}|$ exceeds $\sim 5$ mT, the Zeeman energy becomes larger than the interlayer exchange coupling energy and $\bm{m}_1$ and $\bm{m}_2$ form a scissor state.
The angle of the scissor closes with increasing $|\mu_0 H_\mathrm{ext}|$ and the two layers eventually form the parallel state at $|\mu_0 H_\mathrm{ext}| \sim 20$ mT.
Such change in the magnetization direction can be inferred by the monotonic reduction of $\Delta R_\mathrm{H}$ with increasing $|\mu_0 H_\mathrm{ext}|$ in the range of $\sim 5$ to 20 mT and its saturation at $\sim 20$ mT.


\begin{figure}[bt]
	\begin{tabular}{cc}
		\begin{minipage}{0.5\hsize}
			\includegraphics[scale=0.28]{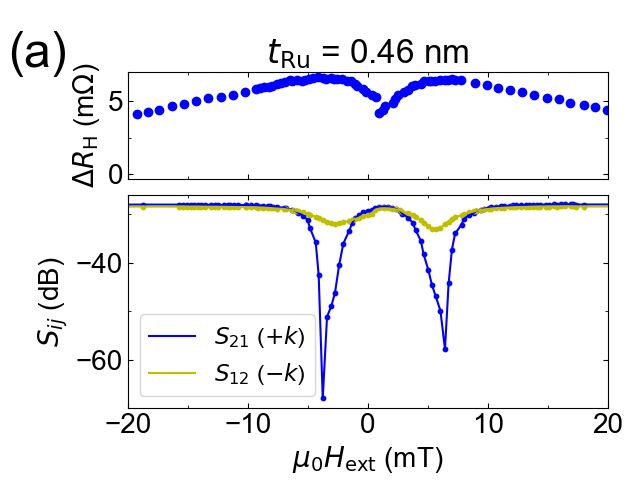}
		\end{minipage}
		\begin{minipage}{0.5\hsize}
			\includegraphics[scale=0.28]{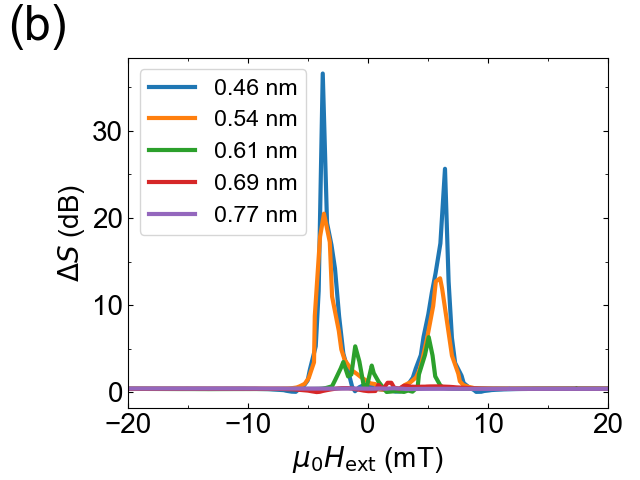}
		\end{minipage}\\
		\begin{minipage}{0.5\hsize}
			\includegraphics[scale=0.28]{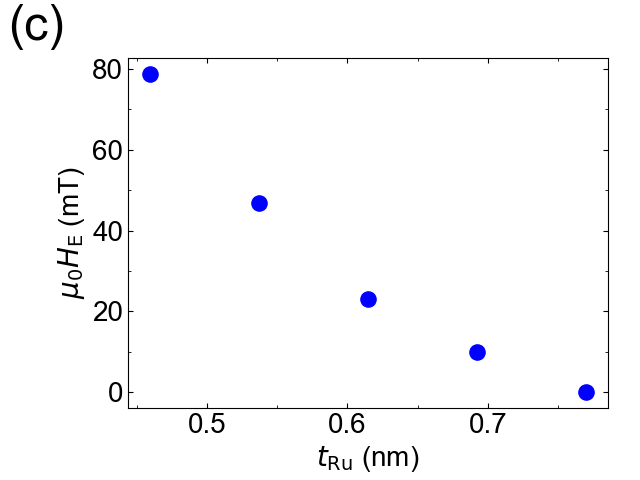}
		\end{minipage}
		\begin{minipage}{0.5\hsize}
			\includegraphics[scale=0.28]{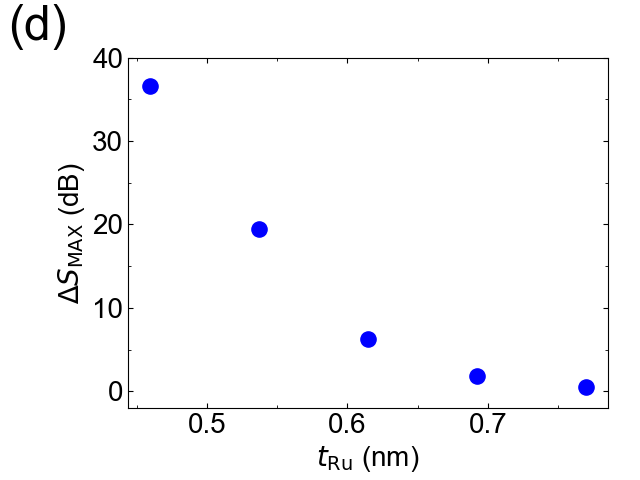}
		\end{minipage}
	\end{tabular}
	\caption{(a) $\Delta R_\mathrm{H}$ (upper panel) and $S_{21}, S_{12}$ (lower panel) dependence on $\mu_{0}H_\mathrm{ext}$ measured for a device with $t_\mathrm{Ru}=0.46$ nm. The magnetic field angle $\varphi_\mathrm{H}$ is set to $\sim - 45^{o}$. (b) Difference in the transmission amplitude ($\Delta S$) when SAW travels along $-x$ and $+x$ plotted as a function of $\mu_{0}H_\mathrm{ext}$ for devices with various $t_\mathrm{Ru}$. (c,d) $t_\mathrm{Ru}$ dependence of $\mu_{0}H_\mathrm{E}$ (c) and $\Delta S_\mathrm{MAX}$ (d).
		\label{fig:nonrec}
	}
\end{figure}

The field at which $\Delta R_\mathrm{H}$ saturates, defined as $\mu_0 H_\mathrm{E}$, represents the anisotropy field of the films.
$H_\mathrm{E}$ includes contributions from the shape (and crystalline) anisotropy field and the interlayer exchange coupling field.
From the PHE measurements, we found that the shape (and crystalline) anisotropy field is less than 1 mT. 
We therefore infer that the enhancement of $H_\mathrm{E}$ with respect to the anisotropy field of single ferromagnetic layer films is caused by the interlayer exchange coupling.
The $t_\mathrm{Ru}$ dependence of $\mu_0H_\mathrm{E}$ is plotted in Fig.~\ref{fig:nonrec}(c). 
The magnitude of the interlayer exchange coupling decreases with increasing $t_\mathrm{Ru}$ and approaches the sum of crystalline and shape anisotropy fields.
For films with ${t}_\mathrm{Ru} \sim 0.77$ nm, we consider the interlayer exchange coupling either vanishes or becomes ferromagnetic.

Propagation of the SAWs across the films is studied using the setup schematically illustrated in Fig.~\ref{fig:setup}(a).
A vector network analyzer (VNA) is connected to the IDTs to measure the scattering matrix of the SAW device.
We define $S_{ij}$ that stands for the amplitude of the scattering matrix of a wave from port $j$ to port $i$.
Representative spectra of $S_{21}$ are shown in Fig.~\ref{fig:setup}(c) and (d) when a film with $t_\mathrm{Ru} = 0.61$ nm is placed in the delay line.
The SAW resonance frequency of the fundamental mode that can be excited by the IDT is $\sim 0.48$ GHz.
We used the 3rd harmonic Rayleigh wave (resonance frequency is $\sim 1.4$ GHz) to excite spin waves in the rectangulaer element, which in turn can cause absorption of the SAWs.
An external magnetic field $H_\mathrm{ext}$ with $\varphi_H \sim -45^\mathrm{o}$ \  is applied during the scattering matrix measurements.
The field angle ($\varphi_H \sim -45^\mathrm{o}$) is chosen since the magneto-elastic coupling takes a maximum when the relative angle between the magnetization and the (SAW-induced) strain is $\sim 45^\mathrm{o} + (90 n^\mathrm{o})$, where $n$ is an integer\cite{dreher2012prb,kawada2021sciadv}.
Maximizing the coupling allows efficient excitation of spin waves.
The amplitude of $H_\mathrm{ext}$ is swept to obtain the magnetic field dependence of $S_{ij}$.

The lower panel of Fig.~\ref{fig:nonrec}(a) shows the $H_\mathrm{ext}$ dependence of the transmission coefficient $S_{12}$ and $S_{21}$ for a device with $t_\mathrm{Ru}=0.46\ \mathrm{nm}$. 
$S_{21}$ ($S_{12}$) represents transmission coefficient of SAWs that travel along the $+x$ ($-x$) direction.
The field is swept from positive to negative $H_\mathrm{ext}$.
The upper panel shows the corresponding $H_\mathrm{ext}$ dependence of $\Delta R_\mathrm{H}$ for the same device.
We found that $S_{12}$ and $S_{21}$ drop at $|\mu_0 H_\mathrm{ext}| \sim 5$ mT.
The field at which the drop occurs is close to that where $\Delta R_\mathrm{H}$ takes a maximum, suggesting that the tilted antiferromagnetically coupled state is essential for the SAW absorption.
Interestingly, the magnitude of the drop depends on the SAW propagation direction.
At $|\mu_0 H_\mathrm{ext}| \sim 5$ mT, $S_{12}$ (transmission coefficient along $-x$) is nearly 30 dB larger than $S_{21}$ (along $+x$), suggesting a significant nonreciprocal SAW transmission is taking place. 
Note that when we reverse the field sweep direction, the relative amplitude of $S_{12}$ and $S_{21}$ changes.

In Fig.~\ref{fig:nonrec}(a), the magnitude of the drop in $S_{12}$ (at $|\mu_{0}H_\mathrm{ext}| \sim 5$ mT) is smaller than that of $S_{21}$ regardless of the sign of magnetic field.
This is in contrast to previous reports, in which the drop at positive $H_\mathrm{ext}$ is larger for one SAW propagation direction and the drop at negative $H_\mathrm{ext}$ is larger for the other SAW propagation \cite{shah2020sciadv,kuss2021prap}.
We infer that this is caused by the difference in the magnetization reversal process.
Here the two exchange coupled ferromagnetic layers are symmetric; both layers are composed of the same material with the same nominal thickness.
Under such circumstance, the relative orientation of the two layers is maintained while $H_\mathrm{ext}$ is varied; see, for example, the sketches in Fig.~\ref{fig:phe}(d).
In contrast, previous studies employed asymmetric structures in which the two coupled ferromagnetic layers are made of different materials and/or different thickness.
The layer with the larger magnetization will always respond to the magnetic field.
Thus when the field direction is reversed, the relative orientation of the two layers changes.
Such change is typically equivalent to reversing the SAW propagation direction, thus causing asymmetry in the absorption for positive and negative fields.  

We characterize the SAW nonreciprocity using a quantity $\Delta S$, defined as  $\Delta S = S_{12} - S_{21}$.
The $H_\mathrm{ext}$ dependence of $\Delta S$ is plotted for devices with different $t_\mathrm{Ru}$ in Fig.~\ref{fig:nonrec}(b).
The peak value of $\Delta S$ decreases as $t_\mathrm{Ru}$ increases, and at $t_\mathrm{Ru} = 0.77$ nm, $\Delta S$ peak nearly disappears.
The maximum $\Delta S$, denoted as $\Delta S_\mathrm{MAX}$, is plotted as a function of $t_\mathrm{Ru}$ in Fig.~\ref{fig:nonrec}(d). 
$\Delta S_\mathrm{MAX}$ increases with decreasing $t_\mathrm{Ru}$, suggesting that the interlayer exchange coupling plays a critical role in defining the nonreciprocity.
The maximum $\Delta S_\mathrm{MAX}$ is obtained for the device with $t_\mathrm{Ru} = 0.46$ nm. 
(Note that the antiferromagnetic interlayer exchange coupling significantly reduces for films with $t_\mathrm{Ru} \lesssim 0.4$ nm, likely due to the non-uniformity of the thin Ru layer.)
For the device with $t_\mathrm{Ru} = 0.46$ nm, the magnitude of nonreciprocity per unit length ($\Delta S_\mathrm{MAX}$ divided by the device length, 1 mm) reaches $\sim 37$ dB/mm, which is larger than that of FeGaB/Al$_{2}$O$_{3}$/FeGaB\cite{shah2020sciadv} (resonance frequency is $\sim 1.4$ GHz) and the same order with CoFeB/Au/NiFe (resonance frequency is $\sim 6.9$ GHz)\cite{kuss2021prap}.
These results therefore show that the strength of the interlayer exchange coupling plays a significant role in defining the nonreciprocity of SAW\cite{nassar2020nrm}.

Previously, the origin of the large nonreciprocity in dipolar coupled FM/NM/FM tri-layers has been associated with the anisotropic dispersion relationship of spin waves due to the dipolar-dipolar coupling\cite{gallardo2019prap, shiota2020prl}. 
The results presented here imply that the interlayer exchange coupling also contributes to the nonreciprocity.
We infer that the coupling may cause non-uniform magnetic structure (e.g. twisted structure) within the FM layer along the film normal, which may increase the anisotropy of the spin wave dispersion relation.
Further studies, including micromagnetic simulations, are required to clarify the exact mechanism that causes the non-reciprocity.




\par
In summary, we have studied the scattering matrix of SAWs that travel across synthetic antiferromagnets (\textit{i.e.}, CoFeB/Ru/CoFeB trilayers). 
We find the SAW transmission across the trilayers depends on its propagation direction.
Such nonreciprocal transport of the SAW manifests itself when the two CoFeB layers are coupled antiferromagnetically.
Interestingly, the nonreciprocity increases with increasing interlayer exchange coupling strength.
When the Ru layer thickness is as thin as 0.46 nm, the nonreciprocity of 1.4 GHz SAW reaches a value of 37 dB/mm, one of the largest values reported thus far.
Our findings may contribute to designing SAW-based isolators and circulators that require significant nonreciprocal wave transmission.
\par
We are grateful to S. Nakatsuji for fruitful discussion. This work was partly supported by JSPS KAKENHI (Grant Nos. 20J20952) from JSPS, and JSR Fellowship, the University of Tokyo.

\bibliography{ref_031622}


\end{document}